# A search for circumstellar dust disks with ADONIS[⋆]

O. Schütz[1], H. Böhnhardt[1], E. Pantin[2], M. Sterzik[3],
S. Els[4], J. Hahn[5], and Th. Henning[1]

[1] Max-Planck-Institut für Astronomie, Königstuhl 17, D-69117 Heidelberg, Germany
[2] DSM/DAPNIA/Service d'Astrophysique, CEA/Saclay, 91191 Gif-sur-Yvette, France
[3] European Southern Observatory, Alonso de Cordova 3107, Casilla 19001, Santiago 19, Chile
[4] Isaac-Newton-Group of Telescopes, Apartado de Correos 321, 38700 Santa Cruz de La Palma, Spain
[5] Institute for Computational Astrophysics, Department of Astronomy and Physics, Saint Mary's University,
Halifax, NS, B3H 3C3, Canada

**Abstract.**
We present results of a coronographic imaging search for circumstellar dust disks with the Adaptive Optics Near Infrared System (ADONIS) at the ESO 3.6 m telescope in La Silla (Chile). 22 candidate stars, known to be orbited by a planet or to show infrared excess radiation, were examined for circumstellar material. In the PSF-subtracted images no clear disk was found. We further determine the detection sensitivities and outline how remaining atmospheric fluctuations still can hamper adaptive optics observations.

## 1. Introduction

Circumstellar disks are a concomitant phenomenon of the formation of stars and planetary systems. Planets are believed to form from planetesimals which in turn are created through coagulation of dust and gas (Beckwith, Henning, & Nakagawa 2000). The final stages of this process remain to be characterised, but observations suggest that the previous protoplanetary accretion disks turn into debris disks after the end of planet formation. Theoretical modelling of debris disks in the presence of a planetary system show structures which can be attributed to its tidal interaction with disk material (Liou & Zook 1999).

Observations of circumstellar matter and searches for new disks bring further understanding on the process of planetary system formation and evolution: in some young and spatially resolved disks, observed with the HST, the existence of asymmetrical structures is attributed to a yet unknown massive body (e.g. in the case of HD 163296, see Grady et al. 2000). Similar conclusions have been drawn from sub-mm observations of dust around Vega and Fomalhaut (Wilner et al. 2002, Holland et al. 2003). Most of the currently known extrasolar planets orbit stars with an age similar to the Sun, where disks already should have disappeared or their emission would be very faint. Therefore, disk detections are often ambiguous. In the case of 55 Cnc, for example, which is a system with three planets, a disk of 40 AU extension and 27° inclination was claimed (Trilling & Brown 1998; Trilling, Brown & Rivkin 2000). Later, from sub-mm observations the disk mass was determined (Jayawardhana et al. 2000). With NICMOS/HST, however, it was not possible to confirm these previous detections (Schneider et al. 2001). For remnant disks an upper lifetime of 400 million years is currently assumed (e.g. Habing et al. 2001).

In this paper we present results from a ground-based, adaptive optics (AO) search for yet unknown circumstellar disks around stars with planets or IR-excess. Some known disks are added for comparison.

## 2. Observation

**Program goal and observing equipment:**

The goal of our observing program was the detection of new disks as well as the search for structures caused by unseen planets in known circumstellar disks. To reach an optimal resolution and brightness contrast we used the ESO Adaptive Optics Near Infrared System (ADONIS) mounted to the 3.6 m telescope at La Silla Observatory (Chile). Typical Strehl ratios are around 0.1 in J-band and 0.3 in H-band. Three observing runs of eight nights in total were allocated to this project between June 2000 and October 2001. ADONIS had been coupled with the near-IR camera SHARPII+ which operates in the J- to K-band. We fur-

**Table 1.** Presentation of our target sample and their stellar parameters. The objects were selected either because of an existing planetary system or observational hints for a disk. See the column *remarks* for a classification of the targets (we do not distinguish between a single planet or a planetary system). Three resolved disks were also included. V-band fluxes are taken from SIMBAD and near-IR fluxes from the 2MASS catalogue. The distances are obtained either via the *Extrasolar Planets Encyclopaedia* (cf. footnote 1) or the cited papers. Where none of this was available the distance is estimated from Hipparcos parallaxes.

(1) HD 141569: Resolved disk (Clampin et al. 2003)
(2) HD 163296: Resolved disk (Grady et al. 2000)
(3) HD 207129: Unresolved disk (Jourdain de Muizon et al. 1999)
(4) HR 4796: Resolved disk (Augereau et al. 1999a)

| Star | d [pc] | Type | V [mag] | J [mag] | H [mag] | K [mag] | Date of Observation | Remarks |
|---|---|---|---|---|---|---|---|---|
| HD 142 | 20.6 | G1IV | 5.70 | 4.69 | 4.65 | 4.47 | Oct. 2001 | planet |
| HD 1237 (= GJ 3021) | 17.6 | G6V | 6.59 | 5.37 | 4.99 | 4.86 | Oct. 2001 | planet |
| HD 4208 | 33.9 | G5V | 7.79 | 6.57 | 6.24 | 6.16 | Oct. 2001 | planet |
| HD 23079 | 34.8 | F8/G0V | 7.1 | 6.03 | 5.81 | 5.71 | Oct. 2001 | planet |
| HD 33636 | 28.7 | G0V | 7.06 | 5.93 | 5.63 | 5.57 | Oct. 2001 | planet |
| HD 52265 | 28.0 | G0V | 6.30 | 5.24 | 5.03 | 4.95 | Oct. 2001 | planet |
| HD 82943 | 27.5 | G0 | 6.54 | 5.51 | 5.25 | 5.11 | Apr. 2001 | planet |
| HD 102647 (= $\beta$ Leo) | 11.1 | A3V | 2.14 | 1.85 | 1.93 | 1.88 | Apr. 2001 | IR-excess |
| HD 134987 | 25.7 | G5V | 6.45 | 5.27 | 5.12 | 4.88 | Apr. 2001 | planet |
| HD 139664 | 17.5 | F5IV-V | 4.64 | 4.02 | 3.73 | 3.80 | Apr. 2001 | IR-excess |
| HD 141569 | 99 | B9.5Ve | 7.0 | 6.87 | 6.86 | 6.82 | Jun. 2000 | known disk [1] |
| HD 155448 | ~600 | B9 | 8.72 | 8.65 | 8.51 | 8.53 | Jun. 2000 | IR-excess |
| HD 158643 (= 51 Oph) | 131 | A0V | 4.81 | 4.90 | 4.71 | 4.30 | Jun. 2000 | PMS star |
| HD 160691 | 15.3 | G3IV-V | 5.15 | 4.16 | 3.72 | 3.68 | Apr. 2001 | planet |
| HD 163296 | 122 | A1Ve | 6.87 | 6.20 | 5.53 | 4.78 | Jun. 2000 | known disk [2] |
| HD 179949 | 27.0 | F8V | 6.25 | 5.30 | 5.10 | 4.94 | Apr. 2001 | planet |
| HD 207129 | 15.6 | G0V | 5.58 | 4.72 | 4.31 | 4.24 | Apr. 2001 | IR-excess [3] |
| HD 217107 | 37.0 | G8IV | 6.18 | 4.95 | 4.76 | 4.54 | Oct. 2001 | planet |
| HD 319139 | ? | K5 | 10.5 | 8.07 | 7.44 | 7.25 | Jun. 2000 | T Tauri |
| HR 4796 | 67 | A0V | 5.78 | 5.78 | 5.79 | 5.77 | Jun. 2000 | known disk [4] |
| HT Lup | ~160 | Ge | 10.4 | 7.57 | 6.87 | 6.48 | Jun. 2000 | T Tauri |
| SAO 185668 | ? | B3 | 9.64 | 8.66 | 8.55 | 8.50 | Jun. 2000 | IR-excess |

ther attached a fully opaque coronographic mask in front of the Lyot (pupil) stop to reject the peak of the PSF. This is necessary to increase the integration time and sensitivity in order to reveal the much fainter emission from circumstellar material. Details on the coronograph and its performance are given in Beuzit et al. (1997). We iteratively centered the star "behind" the mask. To account for the different distances of our targets and thus the varying disk size, coronographic masks with three different diameters were used (0.84″, 1.0″ and 1.4″). These can also be applied to study one object with various sensitivity. Most of the images were acquired with the 1.0″ mask.

**The targets:**

In June 2000 we focused on young stars which were known or suspected to possess a circumstellar disk. In April 2001 we selected stars with known planetary companion, where the survival of a remnant disk is likely. Some targets with IR-excess were also included in this run. The search for remnant disks around recently discovered planetary systems was continued in October 2001. Notes for all targets are shown in Table 1. Further references of the planet parameters can be obtained from the *Extrasolar Planets Encyclopaedia*[1] maintained by Jean Schneider.

**Observing technique:**

Beside of photometric standard stars, all data were acquired with the object centered behind the coronographic mask. We performed the integrations by co-adding several exposures, each single frame taken with the longest reasonable detector integration time, which depending on the targets' brightness was between 0.3 and 60 seconds. The total co-added times are shown in Table 2. Two reference stars (to the best of our knowledge without circumstellar matter) were observed in an alternating sequence with each target for a later subtraction of the point-spread-function (PSF). To reject eventual influences of atmospheric instabilities on the adaptive optics correction, one

---

[1] http://www.obspm.fr/planets

*PSF1-Target-PSF2* cycle was limited to approx. 30 minutes (including all overheads), then the sequence was repeated several further times. Any possible circumstellar matter must be firmly detected in the PSF-subtracted images of each cycle to exclude artifacts caused by atmospheric variations.

From previous ADONIS observations a different response of the adaptive optics correction with regard to the stellar brightness was known. This can result in Strehl ratio variations while it is essential to maintain the conditions between target and PSF star. Therefore, PSF calibration stars were selected which differ not more than 0.3 mag in brightness from the science target, which also have a similar spectral type and which are spatially close to the target (a few degrees).

## 3. Data reduction

**Basic data reduction:**

All raw data were inspected in order to exclude eventual bad images, e.g. structures caused by irregular electronical noise. Data cube clean-up, dark and flatfield correction as well as sky subtraction were performed in a standard way. Flux calibration was achieved via aperture photometry of photometric standard stars.

**PSF subtraction:**

The main reduction step is the optimal subtraction of a reference PSF from the target PSF to unveil the faint circumstellar disk. We use a software developed by E. Pantin (Pantin, Waelkens, & Lagage 2000). For each pair of disk candidate image (Obj) and corresponding PSF the following parameters have to be determined: a scaling factor $R$, sub-pixel shifts $\delta x$ and $\delta y$ between both images in x and y directions, the residual background $Bg$ and the noise standard deviation $\sigma$. These parameters are determined by minimising the penalty functional

$$J = \sum_{pixel} \frac{\left(Obj - \left(shift\left(\frac{PSF}{R}, \delta x, \delta y\right) - Bg\right)\right)^2}{\sigma^2} . \quad (1)$$

The sum is performed over pixels inside a precise mask, not too close to the star, nor too far. Via a software interface showing the resulting image of the optimal PSF subtraction the data reducer checks the numerically obtained fit and can further optimise the subtraction by adjusting some of the parameters. To get an impression of the natural residues also the two PSF stars are subtracted from each other.

## 4. Results

**Reduction artifacts:**

During the reduction process we became aware of several artifacts which despite of a very careful PSF subtraction may occur. These can be characterised as follows:

(a) Diffraction spikes resulting from the mounting of the coronographic mask: these are stripes of varying intensity and

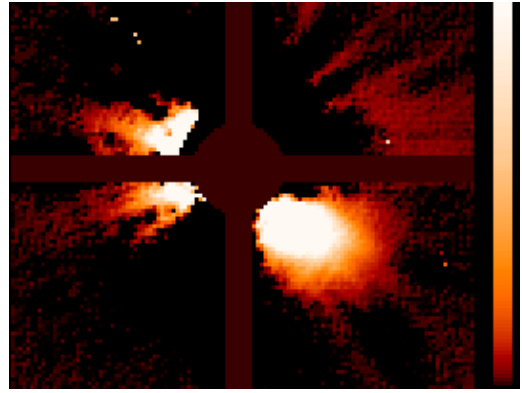

**Fig. 1.** Looking like an edge-on disk around HD 142 (J-band), this is just a very special artifact which we explain in (4c). Intensities therein are scaled logarithmically. Recognisable is the software mask in form of a cross, which is laid over artifacts caused by the coronographic mask. The diameter of the circular mask corresponds to ∼1.2″.

thickness along the wires holding the coronographic mask, i.e. at 0° modulo 90°. They are uncritical and can be hidden with an artificial mask created by our reduction software.

(b) Small rings close to the mask: as a result of an imperfect centering, sickle-shaped structures can appear close to the coronographic mask. Since they are always very close to the mask position, they are easily identified as artifacts and can also be hidden with a software mask.

(c) Diffraction spikes resulting from the support structures of the telescope's secondary mirror: in our special telescope-instrument configuration thin lines at position angles 45° modulo 90° sometimes can be seen, which originate from the mounting of the telescope's secondary mirror. In rare cases and in coincidence with PSF instabilities due to atmospheric variations, these lines may form bright stripes or cause unpredictable results. A very interesting example of this is shown in Fig. 1.

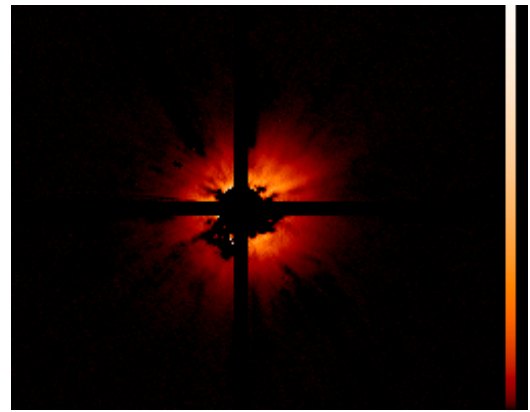

**Fig. 2.** This "face-on disk" around HD 160691 (J-band) was caused by variations in the PSF shape due to atmospheric fluctuations as described in (4d). The intensity is displayed logarithmically. In the software mask the circular part's diameter is ∼1.0″.

| Star | Filter | $t_{exp}$ [min] | $R_{disk}$ [AU] | Sep. ['']  | Limit [mag/$''^2$] | S/N |
|---|---|---|---|---|---|---|
| HD 142 | J | 5 | 50 | 2.5 | 14.2 | 4.1 |
|  |  |  | 75 | 3.7 | 16.7 | 4.3 |
|  |  |  | 100 | 4.9 | 15.0 | 3.5 |
|  | H | 5 | 50 | 2.5 | 14.9 | 3.5 |
|  |  |  | 75 | 3.7 | 17.1 | 3.5 |
|  |  |  | 100 | 4.9 | 16.4 | 2.5 |
| HD 1237 | J | 5 | 50 | 2.9 | 16.0 | 3.6 |
| (= GJ 3021) |  |  | 75 | 4.3 | 16.7 | 2.7 |
|  | H | 7 | 50 | 2.9 | 14.6 | 4.0 |
|  |  |  | 75 | 4.3 | 15.6 | 3.1 |
| HD 4208 | H | 8 | 34 | 1.0 | 13.1 |  |
|  |  |  | 50 | 1.5 | 14.6 | 2.5 |
|  |  |  | 59 | 1.8 | 15.6 |  |
|  |  |  | 75 | 2.2 | 15.9 | 2.7 |
|  |  |  | 85 | 2.5 | 16.4 |  |
|  |  |  | 100 | 3.0 | 16.4 | 2.1 |
|  |  |  | 110 | 3.3 | 17.1 |  |
|  |  |  | 136 | 4.0 | 17.1 |  |
|  |  |  | 161 | 4.8 | 17.7 |  |
| HD 23079 | J | 5 | 50 | 1.5 | 13.5 | 2.3 |
|  |  |  | 75 | 2.2 | 15.2 | 2.6 |
|  |  |  | 100 | 2.9 | 16.0 | 2.4 |
| HD 33636 | J | 5 | 29 | 1.0 | 13.5 |  |
|  |  |  | 50 | 1.8 | 15.0 | 2.6 |
|  |  |  | 72 | 2.5 | 16.7 |  |
|  |  |  | 75 | 2.6 | 16.7 | 3.1 |
|  |  |  | 93 | 3.3 | 17.5 |  |
|  |  |  | 100 | 3.5 | 17.5 | 2.6 |
|  |  |  | 115 | 4.0 | 17.5 |  |
|  |  |  | 136 | 4.8 | 17.5 |  |
| HD 52265 | J | 10 | 50 | 1.8 | 14.2 | 2.6 |
|  |  |  | 75 | 2.7 | 16.0 | 2.6 |
|  |  |  | 100 | 3.6 | 16.0 | 2.3 |
| HD 82943 | H | 2 | 25 | 0.9 | 11.9 | 1.6 |
|  |  |  | 50 | 1.8 | 14.6 | 4.0 |
|  |  |  | 75 | 2.7 | 16.3 | 6.0 |
| HD 102647 | H | 2 | 25 | 2.3 | 13.1 | 3.6 |
| (= β Leo) |  |  | 50 | 4.5 | 14.6 | 5.1 |
| HD 134987 | J | 2 | 25 | 1.0 | 13.2 | 1.8 |
|  |  |  | 50 | 2.0 | 15.9 | 3.8 |
|  |  |  | 75 | 3.0 | 15.9 | 4.9 |
|  | H | 2 | 25 | 1.0 | 12.3 | 1.8 |
|  |  |  | 50 | 2.0 | 13.8 | 3.2 |
|  |  |  | 75 | 3.0 | 15.3 | 3.7 |
| HD 139664 | J | 3 | 25 | 1.5 | 13.6 | 2.8 |
|  |  |  | 50 | 2.9 | 15.1 | 6.6 |
|  |  |  | 75 | 4.3 | 15.9 | 9.1 |
| HD 141569 | H | 5 | 500 | 5.0 | 14.8 | 5.2 |
|  |  |  | 1000 | 10.0 | 15.5 | 3.7 |
|  | $K_s$ | 5 | 500 | 5.0 | 14.5 | 4.3 |
|  |  |  | 1000 | 10.0 | 15.3 | 2.5 |
| HD 155448 | H | 10 | 1000 | 2.0 | - | - |
| HD 158643 | H | 5 | 500 | 3.9 | 13.7 | - |
| (= 51 Oph) |  |  | 1000 | 8.0 | 16.2 | - |

| Star | Filter | $t_{exp}$ [min] | $R_{disk}$ [AU] | Sep. [''] | Limit [mag/$''^2$] | S/N |
|---|---|---|---|---|---|---|
| HD 160691 | J | 2 | 25 | 1.7 | 12.9 | 2.8 |
|  |  |  | 50 | 3.3 | 15.1 | 5.1 |
|  |  |  | 75 | 4.9 | 15.1 | 5.1 |
|  | H | 3 | 25 | 1.7 | 13.1 | 2.8 |
|  |  |  | 50 | 3.3 | 14.6 | 5.8 |
|  |  |  | 75 | 4.9 | 14.6 | 6.5 |
| HD 163296 | H | 6 | 500 | 4.1 | 15.5 | 3.1 |
|  |  |  | 1000 | 8.0 | 16.2 | - |
| HD 179949 | H | 1 | 25 | 1.0 | 12.6 | 2.0 |
|  |  |  | 50 | 1.9 | 15.3 | 3.8 |
|  |  |  | 75 | 2.8 | 16.3 | 5.5 |
| HD 207129 | J | 3 | 25 | 1.6 | 12.9 | 6.5 |
|  |  |  | 50 | 3.2 | 14.4 | 6.5 |
|  |  |  | 75 | 4.8 | 15.1 | 7.8 |
| HD 217107 | J | 5 | 37 | 1.0 | 12.3 |  |
|  |  |  | 50 | 1.4 | 13.8 | 2.4 |
|  |  |  | 65 | 1.8 | 15.2 |  |
|  |  |  | 75 | 2.1 | 15.2 | 3.0 |
|  |  |  | 93 | 2.5 | 16.7 |  |
|  |  |  | 100 | 2.7 | 16.7 | 3.4 |
|  |  |  | 120 | 3.3 | 16.7 |  |
|  |  |  | 148 | 4.0 | 17.5 |  |
|  |  |  | 176 | 4.8 | 17.5 |  |
| HD 319139 | H | 5 | *[1] | 5.0 | 18.0 | 2.9 |
|  |  |  | *[1] | 10.0 | 17.3 | 2.7 |
| HR 4796 [2] | H | 5 | 100 | 1.5 | 12.6 | 1.0 |
|  |  |  | 500 | 7.5 | 17.3 | 2.4 |
|  | $K_s$ | 1 | 100 | 1.5 | 11.5 | 1.4 |
|  |  |  | 500 | 7.5 | 16.5 | 2.0 |
| HT Lup | H | 5 | 500 | 3.1 | 16.4 | 2.6 |
|  |  |  | 1000 | 6.0 | 17.6 | 1.9 |
|  | $K_s$ | 5 | 500 | 3.1 | 15.3 | 2.6 |
|  |  |  | 1000 | 6.0 | 16.5 | 2.3 |
| SAO 185668 | H | 10 | *[1] | 1.0 | - | - |
|  |  |  | *[1] | 5.0 | - | - |

**Table 2.** Integration times and disk magnitude limits. The latter ones were determined with artificial disks of the sizes shown in this table. No disk was found down to the given surface brightness. The reduction routine has a lower performance for stars in clusters and a determination of brightness limits was sometimes not possible. These cases are indicated with '-'. Estimated errors for the given magnitudes are ± 0.2 mag. The disk brightness limit as a function of disk radius was studied in more detail for the targets HD 4208, HD 33636 and HD 217107 with further radii 1.0, 1.8, 2.5, 3.3, 4.0 and 4.8'' (cf. Fig. 3). Disks with radii > 5.0'' are too big for a detection with our software (since they would cover large parts of the detector, which makes them unphysical anyway). The $S/N$-ratio is the $S/N$ per pixel and was calculated as described in Sect. 5.

(1) The distances of HD 319139 and SAO 185668 are unknown.
(2) A faint object at 4.7'' distance to HR 4796 turned out to be a background star, which was also identified by Mouillet et al. (1997).

(d) Variations in the PSF shape due to atmospheric fluctuations: speckles, which result from atmospheric turbulence, can also influence an adaptively corrected PSF for any ground-based observation. As shown in Racine et al. (1999) the speckle noise dominates all other sources of noise (photon, sky and read noise) within the halo. In our data we saw that during unstable atmospheric conditions the PSF shape can even vary within the short time between observation of a target and its corresponding reference star. This naturally results in Strehl ratio fluctuations and a subtraction of these frames would show a circumstellar halo resembling a face-on disk as in Fig. 2.

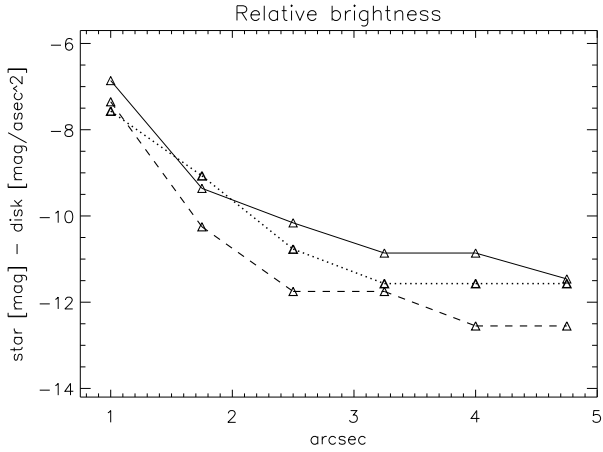

**Fig. 3.** Brightness difference of the star and the disk's detectability limit as function of angular separation. Examples are shown for three targets from Table 2. The dashed line represents HD 217107 (J-band), the dotted line HD 33636 (J-band) and the solid curve is HD 4208 (H-band).

**Disk search and upper limits for disk magnitudes:**

In the PSF-subtracted images of all targets no clear evidence for a disk could be found within the detection sensitivities. In order to estimate for which brightness an eventually existing disk would have been detectable, we carried out a PSF subtraction with artificial disks added to one of the two reference stars. Instead of the target both reference stars were chosen, since we must be sure to use stars without circumstellar matter. As simple-case approximation we used circular, face-on disks with a surface brightness $I(r) = I_0$, i.e. without a radial dependence. $I_0$ was successively reduced until the disk could no longer be revealed by PSF subtraction. The corresponding values are given as mag/arcsec$^2$ in Table 2. Since both the disk radius and $I_0$ were modified, this allows also to conclude on the detectability of disks with other inclinations than face-on. The upper limits determined for an edge-on disk would depend strongly on its orientation and it was not our goal to determine limits for a special, most-favourable case. Realistic disk radii between 25 AU and 100 AU have been chosen for the older and closer stars (11–37 pc) of the runs in April and October 2001. Radii up to 1000 AU were used for the young stars observed in June 2000, which are located at larger distances around 70–600 pc. For HD 4208, HD 33636 and HD 217107 a more detailed analysis of the radial dependence is shown, and the brightness difference of the star and the disk's limit of detectability is plotted in Fig. 3 as function of angular separation.

## 5. Discussion

As can be recognised from Table 2 the detection of circumstellar material with coronography and PSF subtraction still is quite insensitive close to the star, but improves with larger distance. For those targets, where a resolved circumstellar disk is known, we compared the published surface fluxes with the limiting magnitudes from Table 2 and the appropriate disk radii. For HD 141569 ($R_{\text{disk}} \sim 500$ AU) Augereau et al. (1999b) report an H-band surface flux two magnitudes below our detection limit. A difference in brightness applies similar for the $K_s$-band when comparing our values with observations from Boccaletti et al. (2003). Our detection limit for HR 4796 ($R_{\text{disk}} < 100$ AU) is one magnitude brighter than the flux given by Augereau et al. (1999a) for the corresponding separation. A detection of these known disks was marginally missed, but would have been feasible with more integration time.

In order to check whether the atmospheric fluctuations can be identified as the primary cause for the reduced image quality, we now examine the influence of scattered light – both by the atmosphere and the telescope – on our simulated disks. According to Roddier & Roddier (1997) the amount of stellar light scattered by atmospheric turbulence (speckle noise) decreases with the radius $r$ from a star like $r^{-11/3}$. For distances larger than ~3″ scattering by the optical surface roughness of the telescope mirror becomes dominant, which approximately decreases like $r^{-2}$. Our signal $S$ within the simulated disks consists of the radial dependent stellar flux and the model disk with a constant surface brightness:

$$S = F_*(r) + F_{\text{disk}} \qquad (2)$$

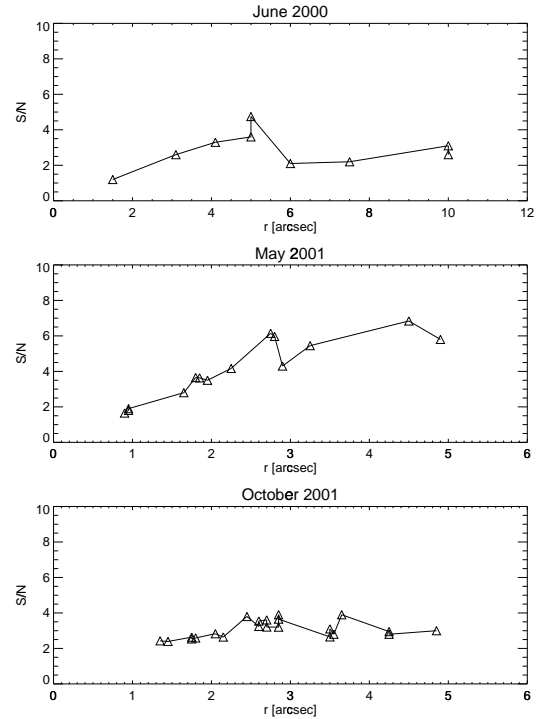

**Fig. 4.** The S/N of the simulated disk detections averaged for each radius over observations in all filters.

The photon noise $N$ is defined by the halo flux of the star at distance $r$:

$$N = \sqrt{F_*(r)} \tag{3}$$

We would expect to have a radial dependence in the $S/N$-ratio going with $r^2$ in case of scattering by the atmosphere and with $r$ in case of scattering by optical surface roughness. The $S/N$ in Table 2 is the $S/N$ per pixel, calculated within a box of $10 \times 10$ pixel at the given radius and averaged over measurements at four position angles which were free of any artifacts. These values are shown in Fig. 4 sorted according to observing run and averaged for each radius over observations in all filters. Results for different runs were not merged in order to account for long-time weather variability between the seasons. From the three plots in Fig. 4 no conclusive results can be seen whether the $S/N$-ratio behaves like $r$ or $r^2$. In the runs of both June 2000 and May 2001 different slopes appear for small resp. larger distances from the star. We speculate that the inner slope could rather be interpreted as a dependence similar to $r^2$ as expected in case of scattering by the atmosphere, while the slope to larger distances shows a rather linear contribution as expected for scattering by the optical surface roughness of the telescope mirror.

## 6. Summary and conclusion

In our data taken with the SHARPII+ / ADONIS system we did not find any clear circumstellar disk. For the known disks the non-detections were caused by insufficient sensitivity and the intrinsic limitations of ground-based AO systems described in this paper.

Many searches for circumstellar material are carried out with AO and coronography. This requires a very stable PSF plus a high Strehl ratio (> 80%, see Sivaramakrishnan 2001), but from any ground-based optical system the PSF stability is limited by atmospheric speckle noise or variations in the width of the PSF caused by atmospheric instabilities. In the latter case, subtracting the PSF of a reference star taken at different atmospheric conditions some minutes before or after the target star, can mimic the existence of a face-on disk.

Dual-imaging techniques like differential polarimetric imaging avoid these problems (Kuhn et al. 2001, Apai et al. 2004): the light from a circumstellar disk is linearly polarised while star light scattered in the earth atmosphere is unpolarised. Two orthogonal polarisation states will be observed simultaneously with a Wollaston prism. Since the atmospherically scattered speckle patterns are unpolarised, the difference image from these two orthogonal polarisations will contain only the polarised light of the real circumstellar disk. Subtraction of a reference PSF – with all the artifacts that may be induced by this – is not required.

*Acknowledgements.* We thank Wolfgang Brandner for his helpful comments and the ESO 3.6 m telescope team for their on-site support. Our study made usage of the SIMBAD database. SE is supported under Marie-Curie Fellowship contract HDPMD-CT-2000-5.